\documentclass[%
 reprint,
superscriptaddress,
 amsmath,amssymb,
 aps,
]{revtex4-2}

\usepackage{graphicx}
\usepackage{dcolumn}
\usepackage{bm}
\usepackage[colorlinks=true,linkcolor=blue,citecolor=blue,urlcolor=blue,]{hyperref}

\begin{document}

\preprint{APS/123-QED}

\title{Numerically stable neural network for simulating Kardar-Parisi-Zhang growth in the presence of uncorrelated and correlated noises}

\author{Tianshu Song}
\affiliation{School of Materials Science and Physics, China University of Mining and Technology, Xuzhou 221116, China}
\affiliation{School of Information and Control Engineering, China University of Mining and Technology, Xuzhou 221116, China}
\author{Hui Xia}%
 \email{hxia@cumt.edu.cn}
\affiliation{School of Materials Science and Physics, China University of Mining and Technology, Xuzhou 221116, China}

\date{\today}

\begin{abstract}
Numerical simulations are essential tools for exploring the dynamic scaling properties of the nonlinear Kadar-Parisi-Zhang (KPZ) equation. Yet the inherent nonlinearity frequently causes numerical divergence within the strong-coupling regime using conventional simulation methods. To sustain the numerical stability, previous works either utilized discrete growth models belonging to the KPZ universality class or modified the original nonlinear term by the designed specified operators. However, recent studies revealed that these strategies could cause abnormal results. Motivated by the above-mentioned facts, we propose a convolutional neural network-based method to simulate the KPZ equation driven by uncorrelated and correlated noises, aiming to overcome the challenge of numerical divergence, and obtaining reliable scaling exponents. We first train the neural network to represent the determinant terms of the KPZ equation in a data-driven manner. Then, we perform simulations for the KPZ equation with various types of temporally and spatially correlated noises. The experimental results demonstrate that our neural network could effectively estimate the scaling exponents eliminating numerical divergence.

\end{abstract}

\maketitle


\section{\label{sec:level1} Introduction\\}


The Kadar-Parisi-Zhang (KPZ) equation \cite{Kardar1986} not only achieves remarkable success in the field of surface growth but also plays a vital role in describing other physical phenomena, such as Bose gas \cite{Diessel2022}, particle transport \cite{Takeuchi2018}, and stirred fluids \cite{Kerstein1992}. The KPZ equation reads
  \begin{equation}
    \frac{\partial h(x, t)}{\partial t}=\nu \nabla^{2} h(x, t)+\frac{\lambda}{2}(\nabla h(x, t))^{2}+ \eta(x, t),
    \label{KPZ}
  \end{equation}
where $h(x,t)$ is the growth height at position $x$ and time $t$, $ \nu $ is the diffusion constant, $\lambda$ is the coefficient of the nonlinear term characterizing lateral growth, and usually $\eta(x,t)$ is Gaussian white noise. The universal behaviour and  scaling exponents of the KPZ equation driven by this type of noise have been fully explored, and satisfactory results have been achieved in terms of both analytical predictions and numerical simulations. However, there are still inconsistent results for the KPZ equation in the presence of  correlated noises. When long-range spatiotemporal correlations are introduced, the noise satisfies
  \begin{equation}
    \left\langle \eta(x, t) \eta\left( x',   t'\right)\right\rangle \sim\left| x- x'\right|^{2 \rho-1}\left| t- t'\right|^{2 \theta-1},
    \label{correlation}
  \end{equation}
where $\rho$ and $\theta$ are spatial and temporal correlation exponents, respectively. The KPZ equation with long-range correlations is challenging to be solved theoretically. Although several theoretical schemes were applied to the correlated growth system \cite{Medina1989,Strack2015,Hanfei1993,Katzav2004,Fedorenko2008,Squizzato2019}, these theoretical predictions have some conflicting  results. 

Meanwhile, previous works \cite{Lam1992,Song2016,Ales2019,Song2021a,Song2021b,Hu2023} performed numerical simulations to explore the scaling properties of the KPZ equation with spatial and temporal correlations.  However, the enormous challenge one must face is that simulating the KPZ equation has annoying numerical instability due to the inherent nonlinearity, which could lead to abnormal growth termination even at the early growth times. 
 In order to avoid this kind of numerical overflow, previous studies typically designed a few special operators \cite{Dasgupta1997,Miranda2008}  to replace the nonlinear term in Eq. (\ref{KPZ}). It was believed that these operators could hold the universality class unchanged. Yet, recent works indicated that this strategy for dealing with the inherent nonlinearity  may notably cause different universality class \cite{Song2021a,Song2021b,Li2021,Hu2023}. Some other works utilized discrete growth models, \it{e.g.} \rm ballistic deposition (BD) model \cite{Lam1992,Song2021a}, to represent the KPZ equation for avoiding numerical divergence. However, recent research revealed that evident discrepancies exist between the BD model and the KPZ equation when long-range correlated noise is introduced \cite{Song2021b}. Consequently, replacing the nonlinear term and utilizing discrete growth models are suboptimal for obtaining convincing results in the presence of long-range correlations. 

\begin{figure*}[t]
  \setlength{\abovecaptionskip}{-0.01cm}

  \centering
     \includegraphics[width=1\linewidth]{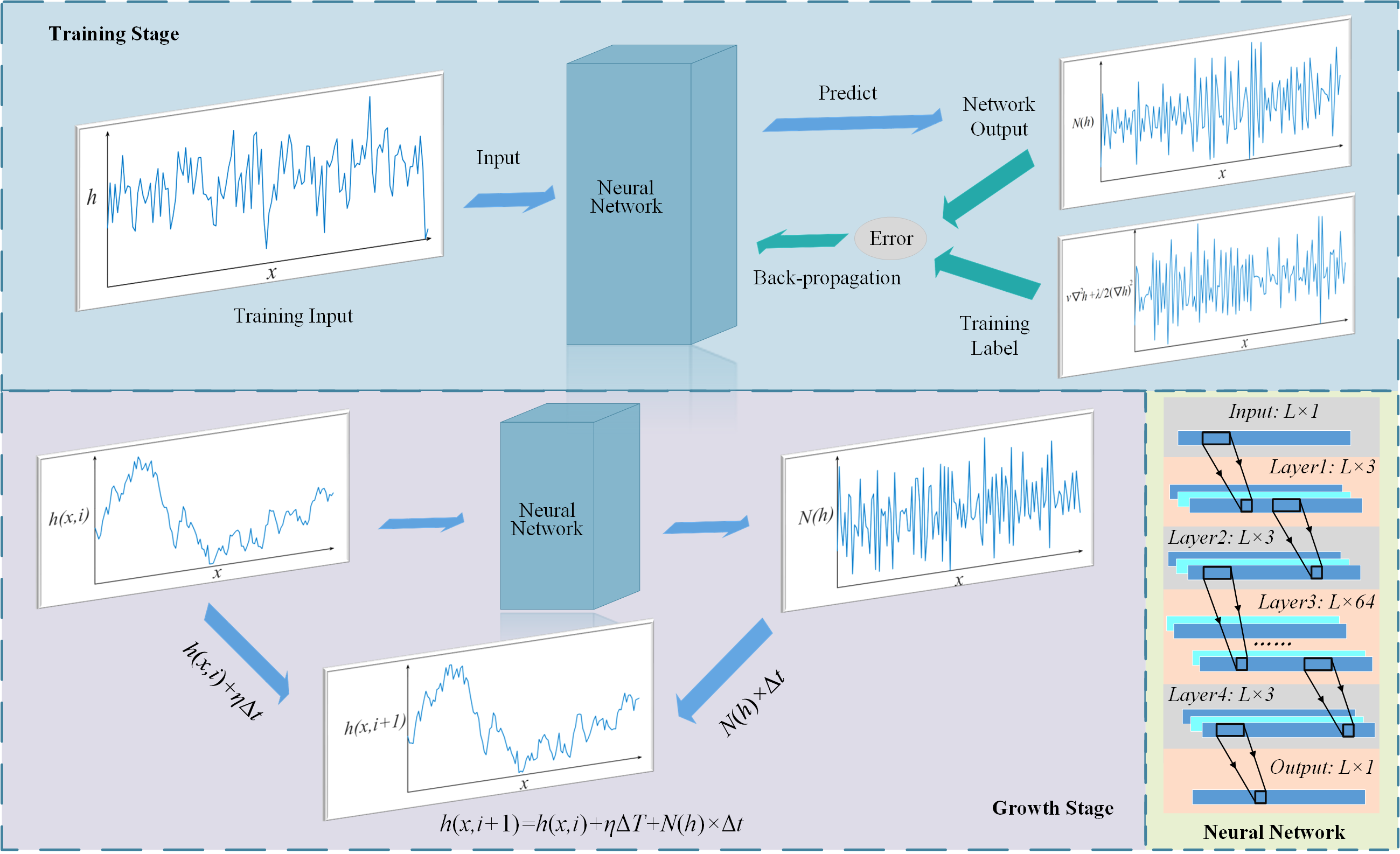}
  \caption{The proposed framework on simulating the stochastic growth system.  }
\label{framework}
\end{figure*}

In recent years, several neural network-based methods have been successively proposed to simulate partial differential equations (PDEs). For example, Raissi \it{et al.} \rm \cite{Raissi2019} proposed a physics-informed neural network to realize the data-driven solution and data-driven discovery of PDEs. Kochkov \it{et al.} \rm \cite{Kochkov2021} adopted a neural network to approximate the Navier-Stokes equations, which achieved 40-80x fold computational speedups for modeling the two-dimensional turbulent flows. Bar-Sinai \it{et al.} \rm \cite{Bar2019} employed a neural network to estimate spatial derivatives to satisfy the PDEs, and the network achieved highly accurate solutions with an unprecedentedly low resolution. The above neural network-based research mainly explored the effectiveness of neural networks for simulating popular PDEs, such as the Navier-Stokes equations, Schrödinger equation, and Burgers equation. However, the related research on dynamic scaling and numerical divergence of the stochastic PDEs is still lacking, which motivates us to investigate this interesting issue, and  the well-known KPZ equation becomes our first choice as a typical representative of various stochastic dynamic systems.

The paper is organized as follows: Firstly, we propose a convolutional neural network-based scheme for simulating the KPZ equation driven by uncorrelated and correlated noises. 
Then, we provide the proposed framework, and implement the detailed training.
Next, we exhibit our numerical simulations and compare them with previous related results. Finally, the corresponding discussions and conclusions are given.

\section{Model and Method}

This section introduces a simple and effective convolutional neural network (CNN)  framework to directly simulate the KPZ equation. Currently, simulating the stochastic PDEs with the neural network is still rarely investigated. Different from the deterministic PDEs (\it{e.g. }\rm Schrödinger and Burgers equations), directly obtaining height field $h(x,t)$ at a specific time and position is impossible because of the unpredictable stochastic disorder.
Therefore, we process the KPZ equation as a discrete-time model, and build a one dimensional convolutional neural network $N$ to represent the first two terms of the right side of Eq. (\ref{KPZ}): 
\begin{equation}
  N(h)=\nu \nabla^{2} h+\frac{\lambda}{2}(\nabla h)^{2}.
  \label{Neural}
\end{equation}
Since the neural network works in a data-driven manner, we need to obtain enough experimental data for training the network. Considering the KPZ equation with Gaussian white noise could be simulated normally with carefully selected parameters, we generate enough training data from simulating  the KPZ equation within the early growth regions before numerical divergence appears.

Based on training data, the network $N$ can commendably satisfy Eq. (\ref{Neural}). Then, we obtain 
 $h(x, t+1)$ by adopting the discretized scheme,
\begin{equation}  
  h(x, t+1) = h(x,t)+[N(h(x, t))+ \eta(x, t))]\Delta t,
  \label{Neural2}
\end{equation}
where $t=1,2,...,T/\Delta t, x=1,2,...,L$. Here, $T$ is the total growth time, $L$ denotes the system size, and $\Delta t$ is the time step.

It should be noted that Eq. (\ref{Neural}) does not contain the noise term, and the network only learns to represent the two determinant terms of the KPZ equation. Therefore, Eq. (\ref{Neural2}) can be adopted to process different uncorrelated and correlated noises. Adopting the proposed method, we can directly change the noise term in Eq. (\ref{Neural2}) to simulate the KPZ equation with both spatially and temporally correlated noise.

The proposed framework has two advantages. Firstly, without directly discretizing the nonlinear term, it can avoid numerical divergence even at a long growth time. Secondly, with the help of convolutional architecture,  the network scheme can directly be adopted to perform simulations with any system size $L$, and without adding additional retraining, it can be further applied to simulate the KPZ system in the presence of long-range temporal or spatial correlation, which is more universal than the traditional simulation methods.

 Figure \ref{framework} shows the proposed framework, which consists of the training and  growth stages. During the training stage,  the network aims to learn the nonlinear mapping of Eq. (\ref{Neural}). The training data of network input $\textbf X$ and output $\textbf Y$  can be obtained by 
  \begin{equation}  
    \left\{\begin{matrix}      
      \vspace{1ex}
      &&  X_{x,t}=h(x,t),\\
      \vspace{1ex}
      &&   Y_{x,t}=\nu [ h( x+ 1,t)-2 h( x,t)+ h( x -  1,t)] \\
      \vspace{1ex}      
      && \quad+(\lambda/8)[ h( x+ 1,t)- h( x- 1,t)]^2,\\
    \end{matrix}\right.
    \label{fd}
  \end{equation}
 with periodic boundary condition. In Eq. (\ref{fd}), $h(x,t)$  follows the traditional finite-difference (FD) scheme,
\begin{equation}  
  \left\{\begin{matrix}
    \vspace{1ex}
    &&h(x,0)=0,\\    
\vspace{1ex}  
    && h( x, t\!+\!1) \!=\!  X_{x,t}+\! [  Y_{x,t}+\!\eta( x, t)]\! \Delta t\!.\\
  \end{matrix}\right.
  \label{fd2}
\end{equation}

After obtaining the training data, we adopt the back-propagation algorithm to train the neural network.  First,  the network  predicts $\textbf Y'_{x,t}$ from input $\textbf X_{x,t}$,
\begin{equation}
  \textbf Y'_{x,t}=N(\textbf X_{x,t}).
  \label{predicts}
\end{equation}
Then, the predictions are sent into the loss function $\mathcal{L}$ for calculating the total error $E$, which reads
\begin{equation}
  E=\mathcal{L}(\textbf  Y'_{x,t},\textbf  Y_{x,t}).
  \label{error}
\end{equation}
Finally, the error is back-propagated \cite{Rumelhart1986} for modifying parameters of $N$:
\begin{equation}
  N_p=N_p-r \frac{\partial E}{\partial N_p},
  \label{bp}
\end{equation}
where $N_p$ denotes the parameters of network $N$, and $r$ is the learning rate.

During the training process, the network satisfies Eq. (\ref{Neural}). Then, at the growth stage, we perform simulations by adopting the commendably trained network. It can be observed from Fig. \ref{framework} that we  send $ h( x, t)$ into the network $N$ and obtain $N(h)$.
Then, by introducing the chosen noise, we obtain $h( x,t+1)$ by the growth rule from Eq. (\ref{Neural2}).


\subsection{Training Details}

During the detailed implementations, we first construct a five-layer neural network, called KPZNet, for simulating the KPZ system, whose structure details  are listed in Tab.  \ref{network}. Then, we obtain the training data $\textbf X_{x,t}$ and $\textbf Y_{x,t}$ by discretizing the KPZ equation with standard Gaussian white noise. The parameters in the KPZ equation we utilized are $ L=128,T=8\times L^{3/2}, \Delta t=0.1, \nu=1,\lambda=4$.  And the initial condition for the growth height is $h(x,0)=0$. At each epoch, we uniformly sample 1/2 number of $t$ for $t<L^{3/2}/\Delta t$ and 1/35  for $t>L^{3/2}/\Delta t$ to reduce the training data number. To increase the diversity of training samples, we repeat this process 1600 times. Finally, we obtain $1.6\times10^8$ training pairs at each epoch for training the neural network.  The loss function consists of two parts. The first part is the  loss function for the predicted $\textbf Y'$ of the neural network: 
\begin{equation}
  E_1\!=\frac{1}{ 2}\sum_{i=1}^{N}\sum_{x=1}^{L}{ (\textbf Y'_{x,t}\!-\!\textbf Y_{x,t})^2}\!+\!\sum_{i=1}^{N}\sum_{x=1}^{L}{ |\textbf Y'_{x,t}\!-\!\textbf Y_{x,t}|},
  \label{error1}
\end{equation}
an the second part of our loss follows the form, 
\begin{equation}
  E_2\!=\frac{1}{ 2}\sum_{i=1}^{N}\sum_{x=1}^{L}{ (\textbf Z'_{x,t}\!-\!\textbf Z_{x,t})^2}\!+\!\sum_{i=1}^{N}\sum_{x=1}^{L}{ |\textbf Z'_{x,t}\!-\!\textbf Z_{x,t}|}.
  \label{error2}
\end{equation}
Here, $\textbf Z'_{x,t},\textbf Z_{x,t}$ are obtained by
\begin{equation}  
  \left\{\begin{matrix}
    \vspace{1.5ex}  
    &&\textbf Z'_{x,t}\!=\!\nu ( \textbf Y'_{x\!+\!1,t}\!+\!\textbf Y'_{x\!-\!1,t}\!-\!2\textbf Y'_{x,t}) \!+\!\frac{\lambda}{2}(\textbf Y'_{x\!+\!1,t}\!-\!\textbf Y'_{x,t})^{2},\\
    &&\textbf Z_{x,t}\!=\!\nu ( \textbf Y_{x\!+\!1,t}\!+\!\textbf Y_{x\!-\!1,t}\!-\!2\textbf Y_{x,t}) \!+\!\frac{\lambda}{2}(\textbf Y_{x\!+\!1,t}\!-\!\textbf Y_{x,t})^{2}.\\
  \end{matrix}\right.
\end{equation}

Thus, the total error $E$ is
\begin{equation}
  E=E_1+E_2.
  \label{total_error}
\end{equation}

By incorporating both absolute error  and squared error, the gradient of the loss function can obtain relatively large values, thereby accelerating the convergence of the network. 
During training, the batch size is set as $N=32$, and the optimizer we utilized is Adam  \cite{Kingma2014}. After each epoch, we test the neural network on the test data $\textbf X^{test}$ and $\textbf Y^{test}$, which are acquired by Eq. (\ref{fd}) with the same parameters during the training stage.   The test error is monitored to avoid overfitting, and the model with the lowest test error is saved.  The initial learning rate is $r=0.003$, and if the test error does not decrease for 2 epochs, the learning rate is multiplied by 0.2. When the learning rate is lower than $5\times 10^{-6}$, we finish the whole training process. 
\begin{table}[t]
  \caption{\label{network}%
  Network Structure of KPZNet.  $N$ and $L$ in the output column means the batch and system sizes, respectively.
  }
  \begin{ruledtabular}
  \begin{tabular}{ccc}
  \textrm{Layer} &\textrm{Kernel} &\textrm{Output}\\
  \textrm{Name} &\textrm{(size, channel, padding)} &\textrm{$batch\times channel\times size$}\\  
  \hline
  \vspace{-2.2mm}\\
  Input &/ &$N\times 1 \times L$\\
  Conv-1 &(3, 3, 1) &$N\times 3 \times L$\\
  Conv-2 &(3, 3, 1) &$N\times 3 \times L$\\
  Conv-3 &(1, 64, 0) &$N\times 64 \times L$\\
  Conv-4 &(1, 3, 0) &$N\times 3 \times L$\\
  Conv-5 &(3, 1, 1) &$N\times 1 \times L$\\
  \end{tabular}
  \end{ruledtabular}
\end{table}
\subsection{Training Results}
\begin{figure}[t]
  \setlength{\abovecaptionskip}{-0.02cm}
  \centering
     \includegraphics[width=0.95\linewidth]{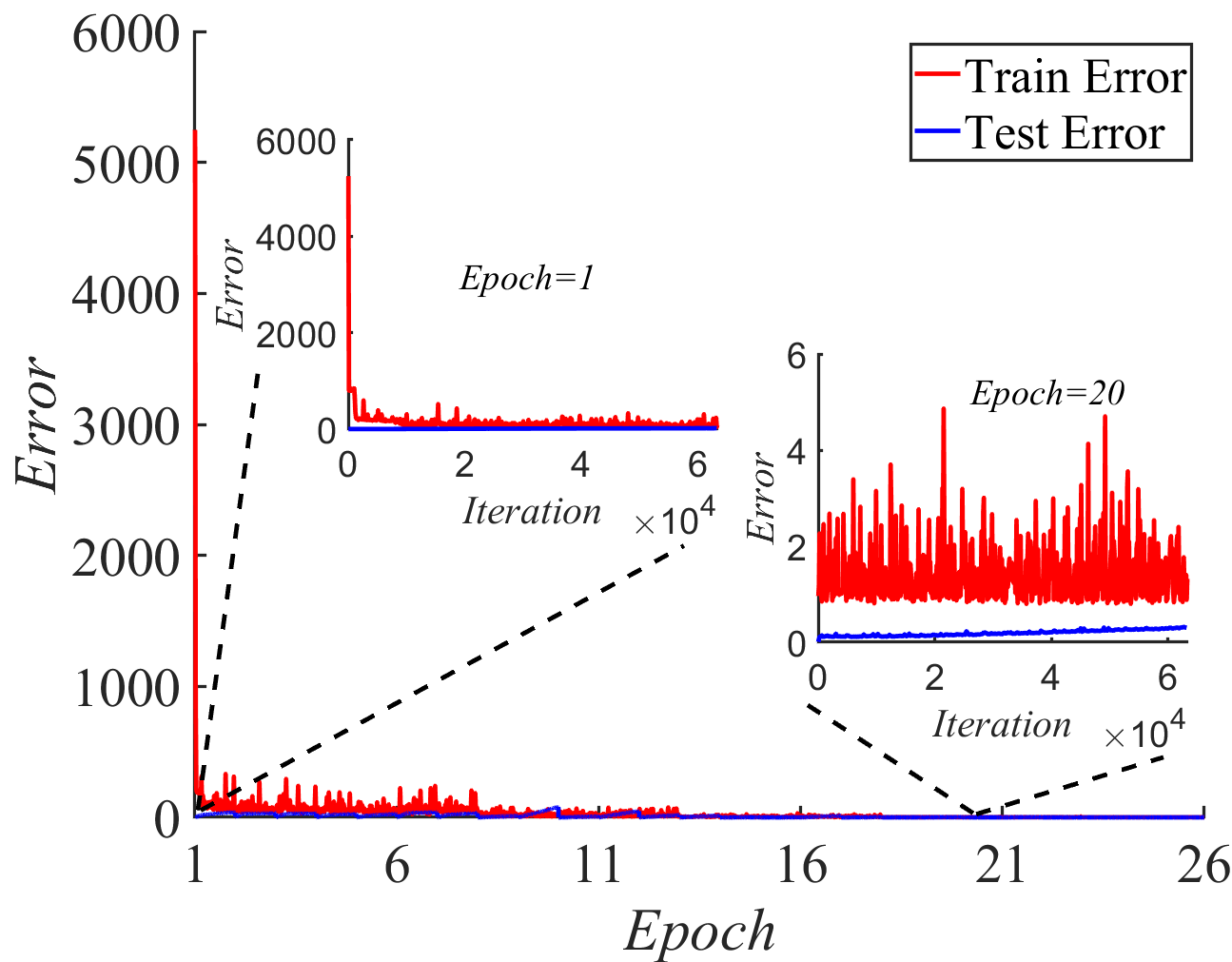}
  \caption{ The errors of the training and test stages. }
  \vspace{-0cm}
  \label{train-loss}
\end{figure}

The errors during training are illustrated in Fig. \ref{train-loss}, and it can be observed that the total errors of both training and test stages quickly decrease and converge to small values, which means the neural network is well-trained to represent the discretized KPZ equation without overfitting.

\begin{figure}[t]
  \setlength{\abovecaptionskip}{-0.02cm}

  \centering
         \includegraphics[width=0.99\linewidth]{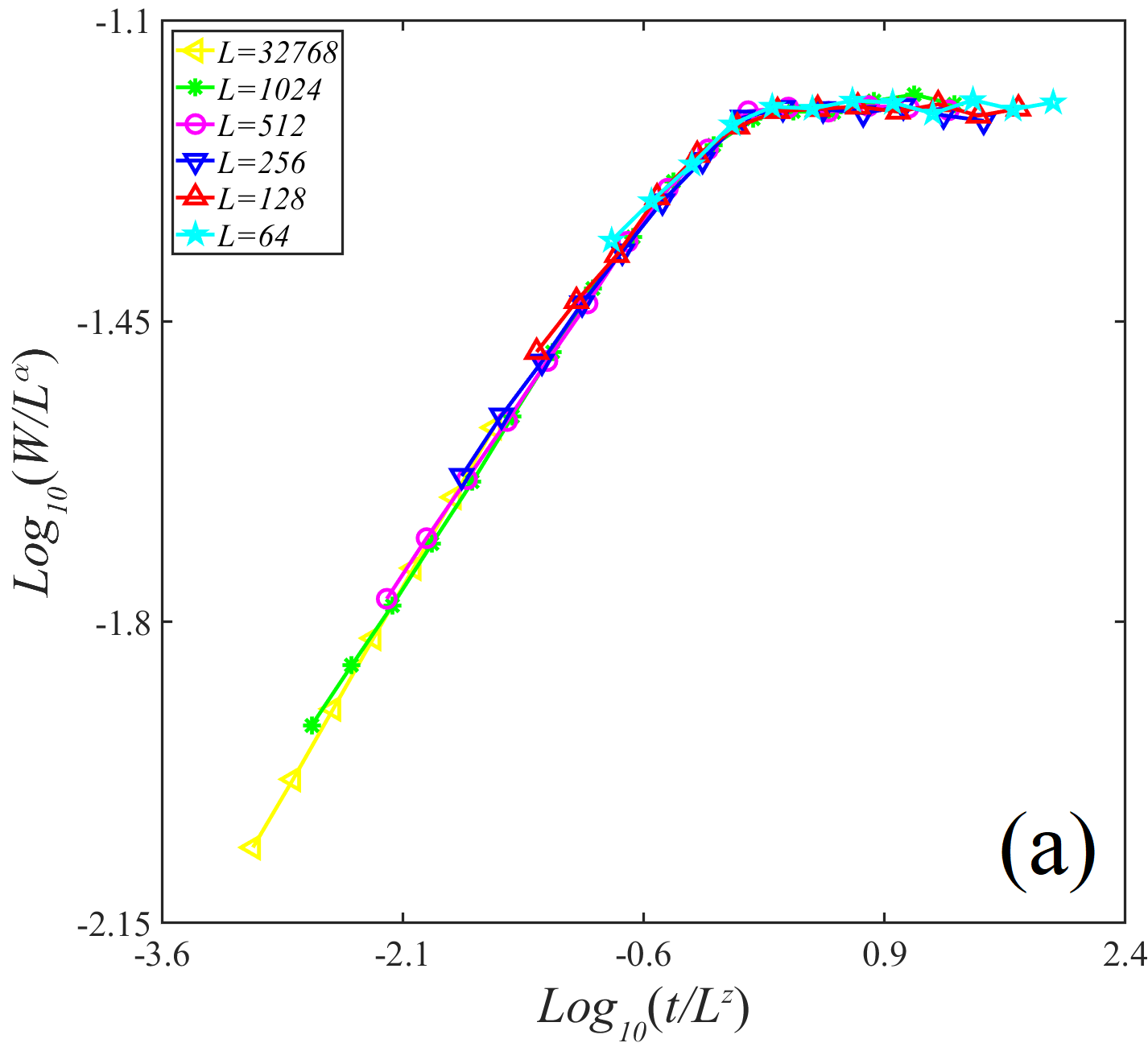}
         \includegraphics[width=1\linewidth]{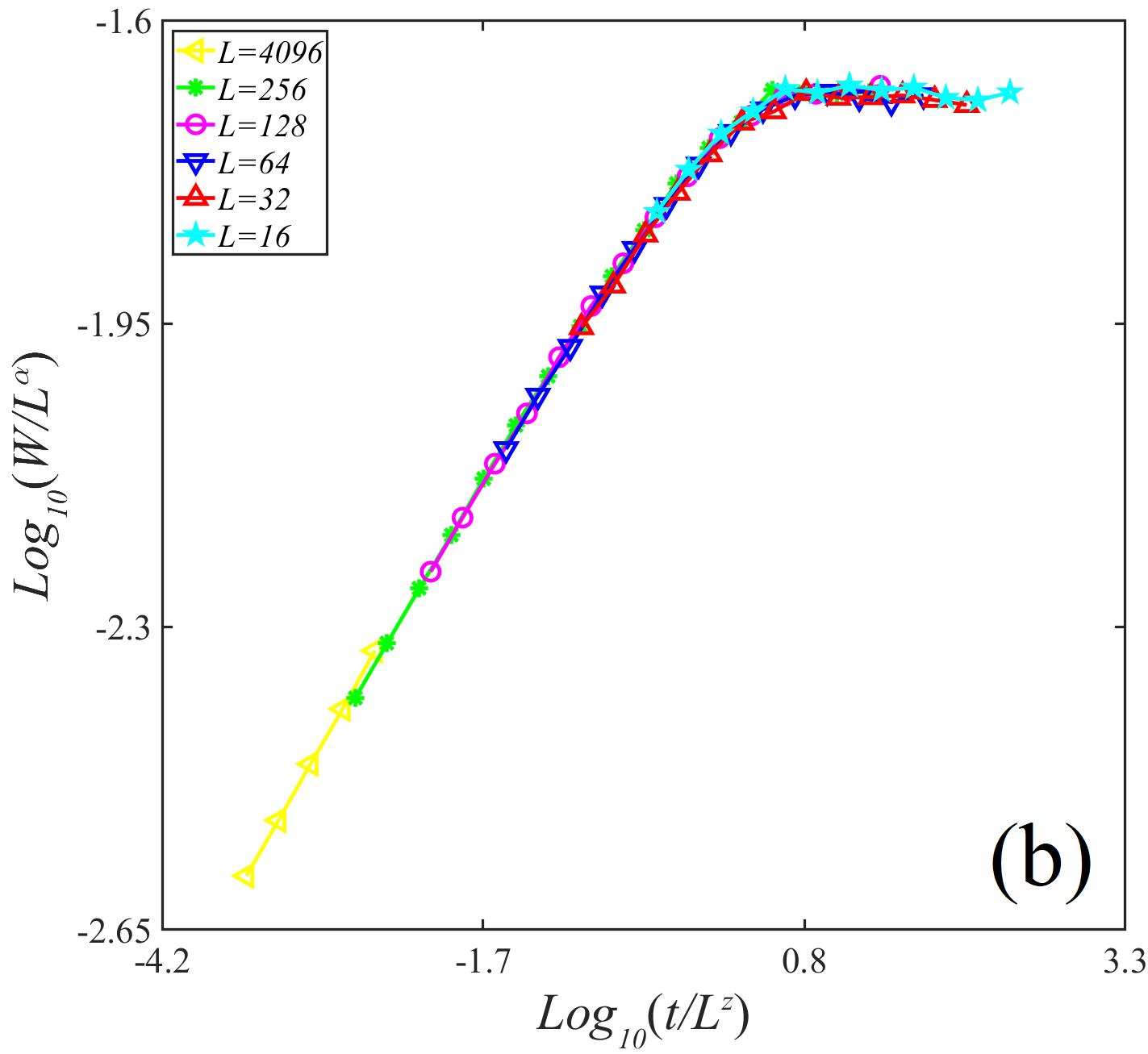}
  \caption{ The log-log plot of the scaled interface width versus the scaled growth time of the growth system driven by Gaussian white noise: (a) KPZ with scaling exponents $\alpha=0.495$ and $z=1.550$, and (b) EW with $\alpha=0.490$ and $z=1.950$. All results have been averaged over 1000 noise realizations.}
\label{Figure3}
\end{figure}

\begin{table}[t]
  \caption{\label{diverge-time}%
   The average divergence time $T_d$ for FD, Cole-Hopf transformation, and the proposed KPZNet with different $\Delta t$ and noise strength $s$.   'None' means that the simulation does not diverge during the whole process ($T=10000$). 
  }
  \begin{ruledtabular}
  \begin{tabular}{ccccc}
  \textrm{$\Delta t$} &\textrm{$s$} &\textrm{$T_d$ (FD)} &\textrm{$T_d$ (Cole-Hopf)}&\textrm{$T_d$ (KPZNet)}\\
  \hline
  \vspace{-2.2mm}\\
  0.01 &1 &None & None &None\\
  0.01 &10 &None & None &None\\
  0.01 &100 &56.8& 165.5 &None\\
  0.01 &1000 &13.4& 28.7 &None\\
  \hline
  0.05 &1 &None & None &None\\
  0.05 &10 &68.7& 1258.6 &None\\
  0.05 &100 &10.0& 38.7 &None\\
  0.05 &1000 &6.0& 18.0 &None\\
  \hline
  0.10 &1 &None & None &None\\
  0.10 &10 &17.7& 149.7 &None\\
  0.10 &100 &7.0& 28.9 &None\\
  0.10 &1000 &5.0& 15.5 &None\\
  \end{tabular}
  \end{ruledtabular}
\end{table}


 To quantitatively describe the scaling properties, we compute the time evolution of interface width $W(L,t)$ representing the fluctuations of height field $h(x,t)$, which satisfies the asymptotic scaling properties $W(L,t)\sim t^{\beta}$ for $t\ll L^{z}$, and $\sim L^{\alpha}$ for $t\gg L^{z}$. Here,  $\beta$, $z$ and $\alpha$ are the growth, dynamic and roughness exponents, respectively.
After training, we adopt the KPZNet to simulate the KPZ equation using Eq. (\ref{Neural2}). 
Figure \ref{Figure3} exhibits the log–log plot of the rescaled $W(L, t) $ versus the rescaled $t $ for the KPZ  growth system driven by Gaussian white noise. The interface width and growth time are rescaled by $W/L^\alpha$ and $t/L^z$, respectively.
It is worth noting that the training parameters need to be reasonably chosen to ensure the simulated system into the true nonlinear KPZ scaling
regime. The obtained results are shown in Fig. \ref{Figure3}(a). 
We find that these scaling exponents estimated from the KPZNet are consistent with the previous numerical results and analytical predictions. Therefore, utilizing the neural network to replace the conventional discretization method is reasonable for dealing with the KPZ equation driven by the uncorrelated noise.

To prove the advantage that the neural network could gain better control over numerical stability, we perform simulations by adopting the KPZNet, the conventional FD method of Eq. (\ref{fd2}), and Cole-Hopf transformation method \cite{Cole1951quasi,Hopf1950partial}. Cole-Hopf transformation is widely utilized to linearize the nonlinear PDEs. Through the transformation,
\begin{equation}
  u\left(x,t\right)=e^{\frac{\lambda}{2\nu}h\left(x,t\right)},
  \label{cole-hopf}
\end{equation}
we obtain the field $u\left(x,t\right)$ from the KPZ equation: 
\begin{equation}
  \frac{\partial u\left(x,t\right)}{\partial t}=\nu\nabla^2u\left(x,t\right)+\frac{\lambda}{2\nu}u\left(x,t\right)\eta\left(x,t\right).
  \label{cole-hopf2}
\end{equation}
The discretized scheme of Eq. (\ref{cole-hopf2}) can be written as follows,
\begin{equation}  
  \left\{\begin{matrix}
    \vspace{1.5ex} 
    && u(x,0)=1,\\\vspace{0.5ex} 
    &&u\left(x,t+\Delta t\right)=[u\left(x+1,t\right)+u\left(x-1,t\right)-2u\left(x,t\right)\\
    \vspace{1.5ex} 
    &&+\frac{\lambda}{2\nu}u\left(x,t\right)\eta\left(x,t\right)]\Delta t +u\left(x,t\right), \\
    &&h\left(x,t+1\right)=\frac{2\nu}{\lambda}ln[u\left(x,t+1\right)].\\
  \end{matrix}\right.
  \label{fd-cole-hopf2}
\end{equation}

Then, we make numerical experiments with different time steps $\Delta t=0.01,0.05,0.1$,  the chosen noise $\eta=s\times  N(0,1)$ with different noise strength $s=1,10,100,1000$, and $ N(0,1)$ is the standard Gaussian white noise.  
We totally simulate $10000$ steps and record the divergence time $T_d$. For each parameter, we simulate $10$ independent runs and estimate the average $T_d$, which is shown in Tab. \ref{diverge-time}.    \color{black}  We find that the divergence phenomenon could be effectively avoided by adopting KPZNet. It is worth mentioning that the proposed KPZNet can obtain effective scaling exponents with  $s=1000 $ and $ \Delta t=0.05$, which is also consistent with the previous results \cite{Song2021b,Hu2023}.

Here, we provide the possible reasons for the numerical stability of the proposed KPZNet. Firstly, the KPZNet has a larger receptive field compared with the FD method.  Considering the architecture of the proposed five-layer convolutional network, the prediction height  $h\left(x,t+1\right)$ can be obtained from the seven positions for the (1+1)-dimensional case, including $h\left(x-3,t\right),\ h\left(x-2,t\right)$, $h\left(x-1,t\right), h\left(x,t\right)$, $h\left(x+1,t\right)$, $h\left(x+2,t\right), h\left(x+3,t\right)$. The KPZNet describes the time evolution of the KPZ equation in its own form by incorporating more data points. By increasing the number of data points, it is possible to sample the variations for capturing additional details and features. This is particularly crucial for equations characterized by rapid changes or local peculiarities. Then, the simulation results can exhibit improved smoothness in the spatial region, which helps reduce numerical errors and oscillations, ultimately enhancing the numerical stability.

In addition to numerical stability, computational efficiency is another crucial factor in performing simulations. We evaluate the performance of the FD and Cole-Hopf transformation methods, and the proposed KPZNet on two different GPU devices: the NVIDIA TitanXP (with Intel 2×E5-2630v4 CPUs) and the NVIDIA RTX3090 (with an Intel Core-i7 13700KF CPU).  More specifically, our simulations involve the KPZ driven by Gaussian white noise with $L=1024$ and $t=10^5$. For the sake of reliable results, 10 independent ensemble averages are used. The execution times of three methods are listed in Tab. \ref{efficiency}.
 It is easy to find that the FD and Cole-Hopf transformation methods take similar amounts of run times, which are faster than the proposed KPZNet scheme. Thus, exploring how to improve the computational efficiency using neural networks deserves further investigation.

\begin{table}[t]
  \caption{\label{efficiency}%
  The execution time (seconds) for FD, Cole-Hopf transformation and the proposed KPZNet. 
  }
  \begin{ruledtabular}
  \begin{tabular}{cccc}
     \textrm{GPU}  &\textrm{ FD} &\textrm{  Cole-Hopf}&\textrm{ KPZNet}\\
  \hline
  \vspace{-2.2mm}\\
    TitanXP  & 51.4s & 53.2s & 110.1s\\
    RTX3090 & 20.0s & 20.9s &39.9s\\  
  \end{tabular}
  \end{ruledtabular}
\end{table}

\section{Numerical results in (1+1) dimensions}
Since the KPZNet is trained as an end-to-end paradigm, and the training data are obtained from directly simulating the KPZ equation, one might worry that it just learns to map $ h(x,t)$ obtained from the KPZ universality class. However, we desire that the KPZNet can be generalized to process the height field with different distributions. In fact, when $\Delta t$ in Eq. (\ref{fd2}) is small enough, the obtained results could crossover to those of the Edwards-Wilkinson (EW) class \cite{Edwards1982}, which is equivalent to reducing the nonlinear coefficient $\lambda$ to a critical threshold. 
Recently, another alternative protocol was introduced by a small number of the controlled Fourier modes for the crossover from strong nonlinear to weak coupling in the KPZ growth process \cite{Priyanka2021}.
Then we use a small $\Delta t$ to perform simulations based on the KPZNet, and the corresponding results obtained are illustrated in Fig. \ref{Figure3}(b). In this case, the scaling exponents obtained from KPZNet could be well consistent with the exact values of the EW universality class. This experiment proves that the KPZNet can be generalized to process $h(x,t)$ with other distributions of the training data, indicating that the KPZNet has grasped the growth rule behind the training data.



\begin{figure}[t]
  \setlength{\abovecaptionskip}{-0.02cm}
  \centering
     \includegraphics[width=1\linewidth]{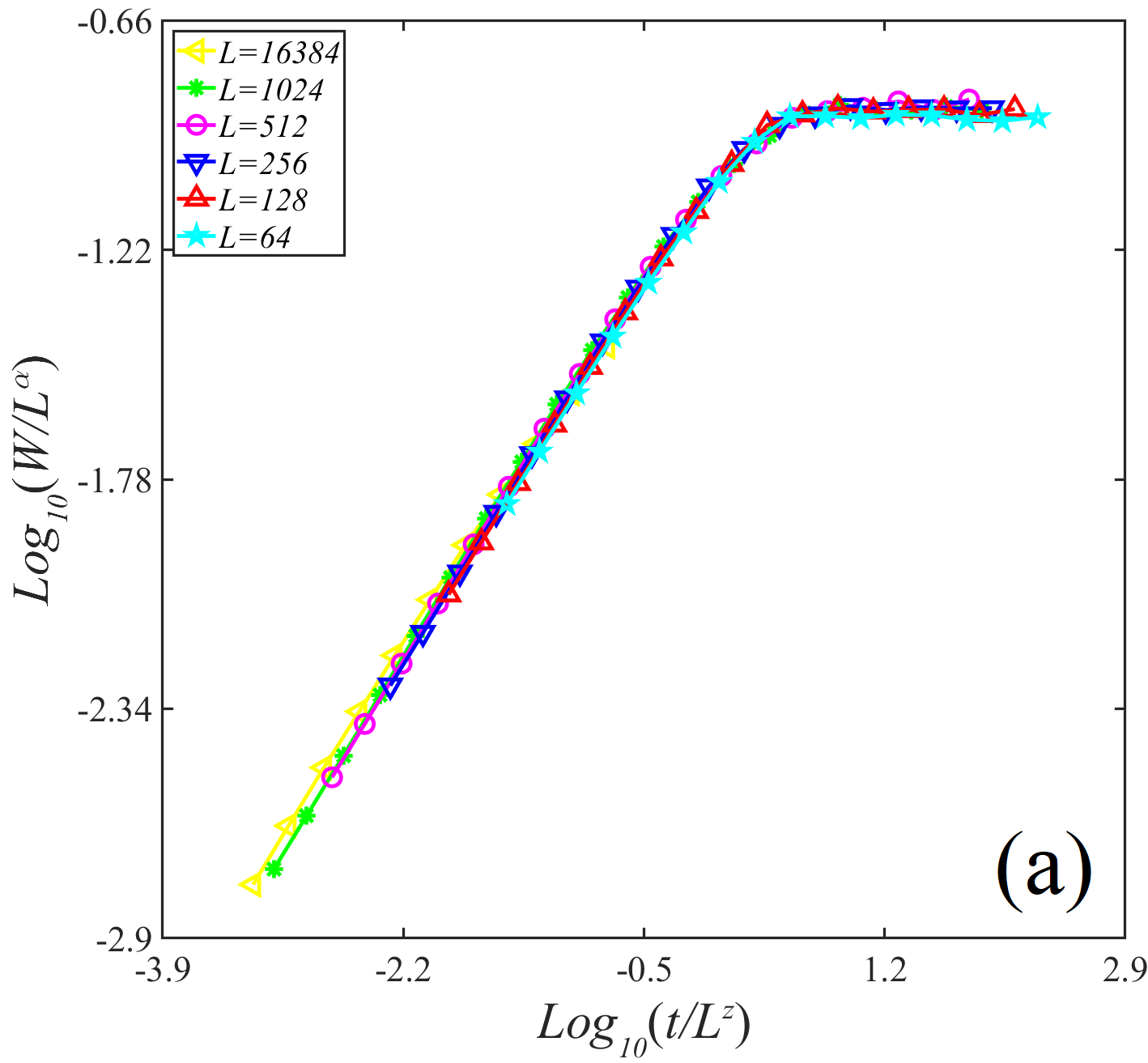}
    \includegraphics[width=1\linewidth]{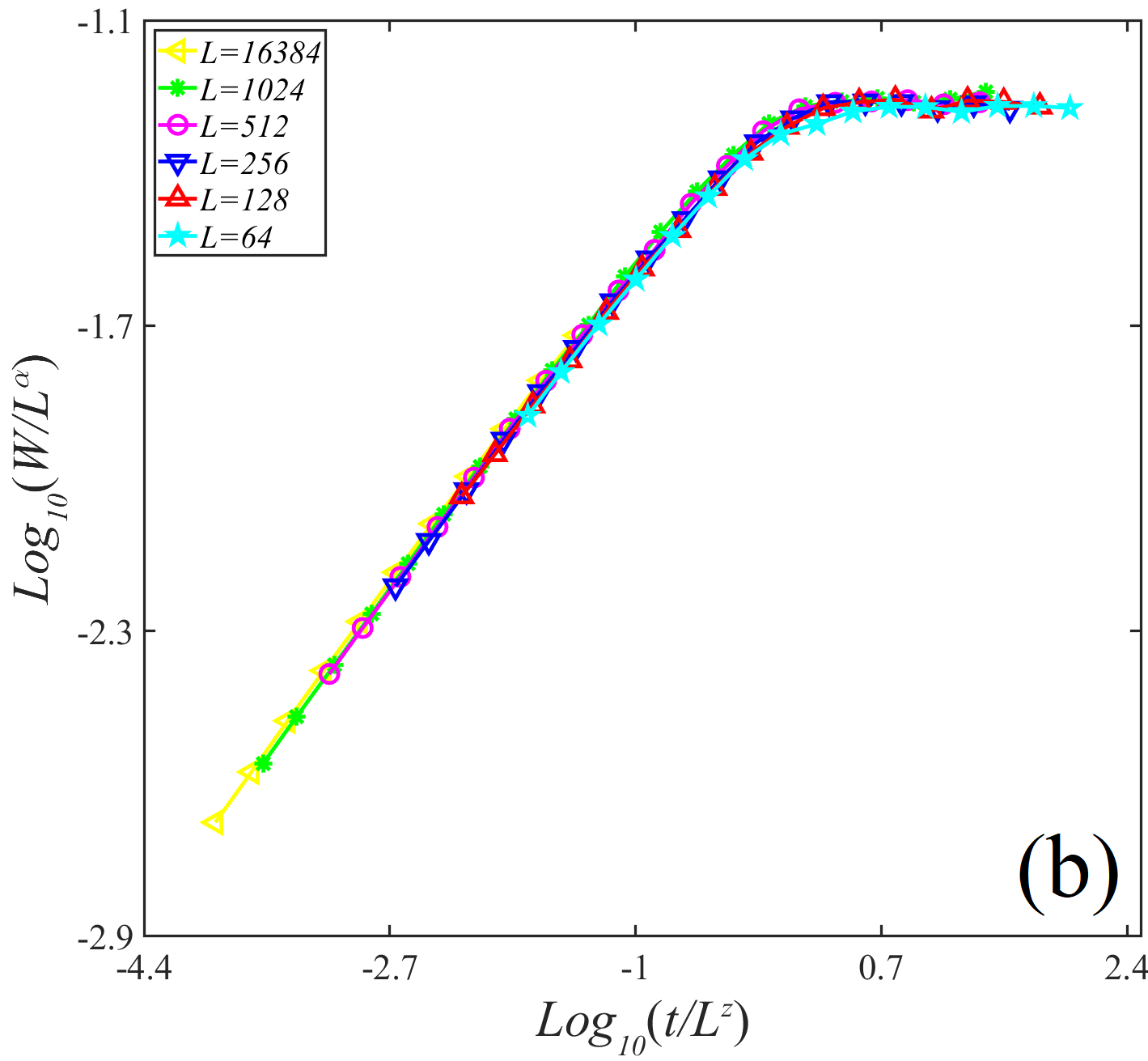}
  \caption{ The collapse results of scaled interface width versus the scaled growth time from (1+1)-dimensional KPZ driven by (a) long-range temporal correlations ($\theta=0.30$) with $\alpha=0.745$ and  $z=1.370$, and long-range spatial correlations ($\rho=0.30$) with $\alpha=0.610$ and $z=1.520$. All data were averaged over 500 independent noise realizations.}
\label{Figure4}
\end{figure}

To further study the numerical stability and dynamic scaling properties in dealing with the correlated KPZ system, we adopt the KPZNet to simulate the KPZ equation driven by temporally correlated noise. The method we utilized for generating the long-range correlations is the fast fractional Gaussian noise (FFGN) algorithm \cite{Mandelbrot1969}. In order to study long-range temporal correlations in the KPZ growth, previous numerical investigations usually adopted discrete growth models with the discrete noise $[0, 1]$  or  modified the nonlinear term of the KPZ equation. In order to avoid possible deviation from the true KPZ class, the proposed KPZNet directly adopts Eq.(\ref{Neural2}) to perform numerical simulations. 
The scaling results of the temporally correlated KPZ system are obtained, as shown in Fig. \ref{Figure4}(a). Here, $\theta=0.30$ is chosen as the typical representative of long-range temporal correlations. Finally, we perform simulations on the KPZ with spatially correlated noise, and the simulating results ($\rho=0.30$) are illustrated in Fig. \ref{Figure4}(b).  Whether for temporally or  spatially correlated KPZ systems, the scaling exponents are estimated correspondingly based on good data collapses.

Numerical divergence in simulating the correlated KPZ system happens easily using a common simulation method without specified treatment. In contrast, Figures. \ref{Figure4}(a)-(b) exhibit that, by utilizing the neural network, the proposed method can grow to the saturation stage without numerical divergence, and the scaling exponents obtained from correlated noise are consistent with that of previous results within the error range \cite{Song2021b,Hu2023}. It should be noted that the desired results are achieved by directly adopting Eq.(\ref{Neural2}) without any extra process (\it{e.g. }\rm changing the amplitude of noise or other parameters),  indicating the high availability of the KPZNet for temporally and spatially correlated noises. 

All the above experiments show that the proposed KPZNet achieves effective scaling exponents for the (1+1)-KPZ growth driven by uncorrelated and correlated noises. The neural network could reveal the main rules  behind the training data. Without additional adjustments, it could also be applied to deal with the linear and nonlinear growth system driven by various types of noises. Thus, adopting the proposed numerical stable neural network is a promising approach to perform various types of stochastic dynamic systems numerically.  In addition to grasping the knowledge of the KPZ equation, we believe that the diversity of training data significantly influences the generalization ability. The model predicts results at each x by considering several positions around the local region. The surface of KPZ driven with Gaussian noise exhibits a more complex and diverse structure in local regions. As a result, the KPZNet trained with Gaussian noise demonstrates strong generalization ability.

\section{Numerical experiments in (2+1) dimensions}
\begin{table}[t]
  \caption{\label{network2d}%
  Network Structure of KPZNet2d.  $N,L$ in the output size respectively means batch size and system size.
  }
  \begin{ruledtabular}
  \begin{tabular}{ccc}
     \textrm{Layer} & \textrm{Kernel} & \textrm{Output}\\
     \textrm{Name} & \textrm{(size, channel, padding)} & \textrm{$batch\times channel\times size$}\\  
  \hline
  \vspace{-2.2mm}\\
   Input & / & $N\times 1 \times (L \times L)$\\
   Conv-1 & ($3\times 3$, 8, 1) & $N\times 3 \times (L \times L)$\\
   Conv-2 & ($3\times 3$, 16, 1) & $N\times 3 \times (L \times L)$\\
   Conv-3 & ($3\times 3$, 32, 1) & $N\times 64 \times (L \times L)$\\
   Conv-4 & ($3\times 3$, 64, 1) & $N\times 3 \times (L \times L)$\\
   Conv-5 & ($3\times 3$, 1, 1) & $N\times 1 \times (L \times L)$\\
  \end{tabular}
  \end{ruledtabular}
\end{table}
  In this section, we generalize numerical experiments to the (2+1)-dimensional case. First, we design a (2+1)-dimensional convolutional neural network of the KPZ system (denoted as KPZNet2d), whose structure is listed in Table \ref{network2d}.

  Then, we generate the training data of $\textbf X,\textbf Y $:
  \begin{equation}  
    \left\{\begin{matrix}      
      \vspace{1.5ex}
      && \textbf X_{x,y,t}=h(x,y,t),\\
      \vspace{0.5ex}
      &&  \textbf Y_{x,y,t}=\nu [ h( x+ 1,y,t)-2 h( x,y,t)+ h( x -  1,y,t)\\
      \vspace{0.5ex}
      &&+h( x,y+1,t)-2 h( x,y,t)+ h( x,y-1,t)] \\
      \vspace{0.5ex}      
      && \quad+(\lambda/8)[ h( x+1,y,t)- h( x- 1,y,t)]^2\\
      && \quad+(\lambda/8)[ h( x,y+1,t)- h( x,y-1,t)]^2,\\

    \end{matrix}\right.
    \label{data2d}
  \end{equation}
where $t=1,2,...,T/\Delta t$, and periodic boundary condition is used. In Eq. (\ref{data2d}), $h(x,y,t)$  follows the discrete version using the traditional FD method:

\begin{equation}  
  \left\{\begin{matrix}
    \vspace{1.5ex}
    &&h(x,y,1)=0,\\      
    && h( x,y, t\!+\!1) \!=\!  h( x,y,t)\!+\!\langle\nu [ h( x\!+\! 1,\!y,\!t)\!+\! h( x \!-\! 1,\!y,\!t)\\
    \vspace{0.5ex}
    &&\!+\!h( x,y\!+\!1,t)\!+\! h( x,y\!-\!1,t)-\!4 h( x,y,t)]\\
    \vspace{0.5ex}      
    &&  \!+\!\frac{\lambda}{8}[ h( x\!+\! 1,y,t)\!-\! h( x\!-\! 1,y,t)]^2\!+\!\frac{\lambda}{8}[ h( x,y\!+\!1,t)\\
    \vspace{0.5ex} 
    && \!-\! h( x,y\!-\!1,t)]^2+\eta(x,y,t)\rangle\Delta t. \\
    
  \end{matrix}\right.
  \label{fd2d}
\end{equation}

Here, the chosen parameters for the (2+1)-dimensional case are $ T=8\times L^{3/2}, \Delta t=0.5, \nu=0.25,\lambda=4$.  We uniformly sample 1/5 number of $t$ for $t<L^{3/2}/\Delta t$ and 1/35  for $t>L^{3/2}/\Delta t$ to reduce the training data number. 

Next, we train the model with $\textbf X,\textbf Y $ and the loss function is:  

\begin{equation}  
  \left\{\begin{matrix}
    \vspace{0.5ex} 
    &&E_1=\frac{1}{ 2}\sum_{i=1}^{N}\sum_{x=1}^{L}\sum_{y=1}^{L}{ (\textbf Y'_{x,y,t}-\textbf Y_{x,y,t})^2}\\
    \vspace{1.5ex} 
    &&+\sum_{i=1}^{N}\sum_{x=1}^{L}\sum_{y=1}^{L}{ |\textbf Y'_{x,y,t}-\textbf Y_{x,y,t}|},\\\vspace{0.5ex} 
    
    &&E_2=\frac{1}{ 2}\sum_{i=1}^{N}\sum_{x=1}^{L}\sum_{y=1}^{L}{ (\textbf Z'_{x,y,t}-\textbf Z_{x,y,t})^2}\\
    \vspace{1.5ex} 
    &&+\sum_{i=1}^{N}\sum_{x=1}^{L}\sum_{y=1}^{L}{ |\textbf Z'_{x,y,t}-\textbf Z_{x,y,t}|},\\   
    \vspace{0.5ex} 
    &&\textbf Z'_{x,y,t}=\nu ( \textbf Y'_{x+1,y,t}+\textbf Y'_{x-1,y,t}-2\textbf Y'_{x,y,t}\\ \vspace{0.5ex} 
    &&+\textbf Y'_{x,y+1,t}+\textbf Y'_{x,y-1,t}-2\textbf Y'_{x,y,t}) \\ \vspace{0.5ex} 
    &&    +\lambda/8(\textbf Y'_{x+1,y,t}-\textbf Y'_{x-1,y,t})^{2}\\
    \vspace{1.5ex} 
    &&    +\lambda/8(\textbf Y'_{x,y+1,t}-\textbf Y'_{x,y-1,t})^{2},\\\vspace{0.5ex} 

    &&\textbf Z_{x,y,t}=\nu ( \textbf Y_{x+1,y,t}+\textbf Y_{x-1,y,t}-2\textbf Y_{x,y,t}\\\vspace{0.5ex} 
    &&+\textbf Y_{x,y+1,t}+\textbf Y_{x,y-1,t}-2\textbf Y_{x,y,t}) \\\vspace{0.5ex} 
    &&    +\lambda/8(\textbf Y_{x+1,y,t}-\textbf Y_{x-1,y,t})^{2}\\
    \vspace{1.5ex} 
    &&    +\lambda/8(\textbf Y_{x,y+1,t}-\textbf Y_{x,y-1,t})^{2},\\

    &&E=E_1+E_2.
  \end{matrix}\right.
\end{equation}

\begin{figure}[t]
  \setlength{\abovecaptionskip}{-0.02cm}
  \centering
     \includegraphics[width=1\linewidth]{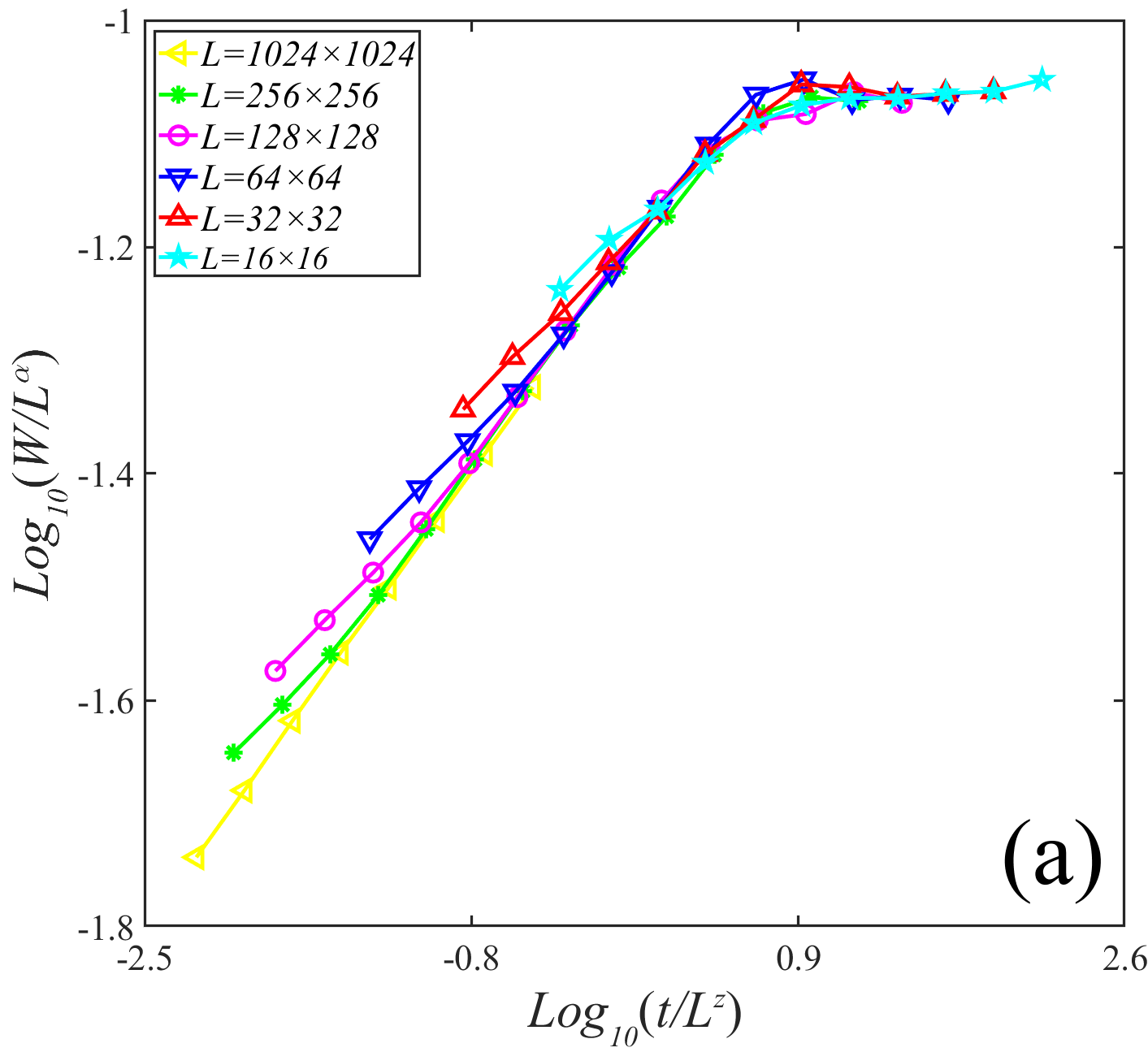}
    \includegraphics[width=1\linewidth]{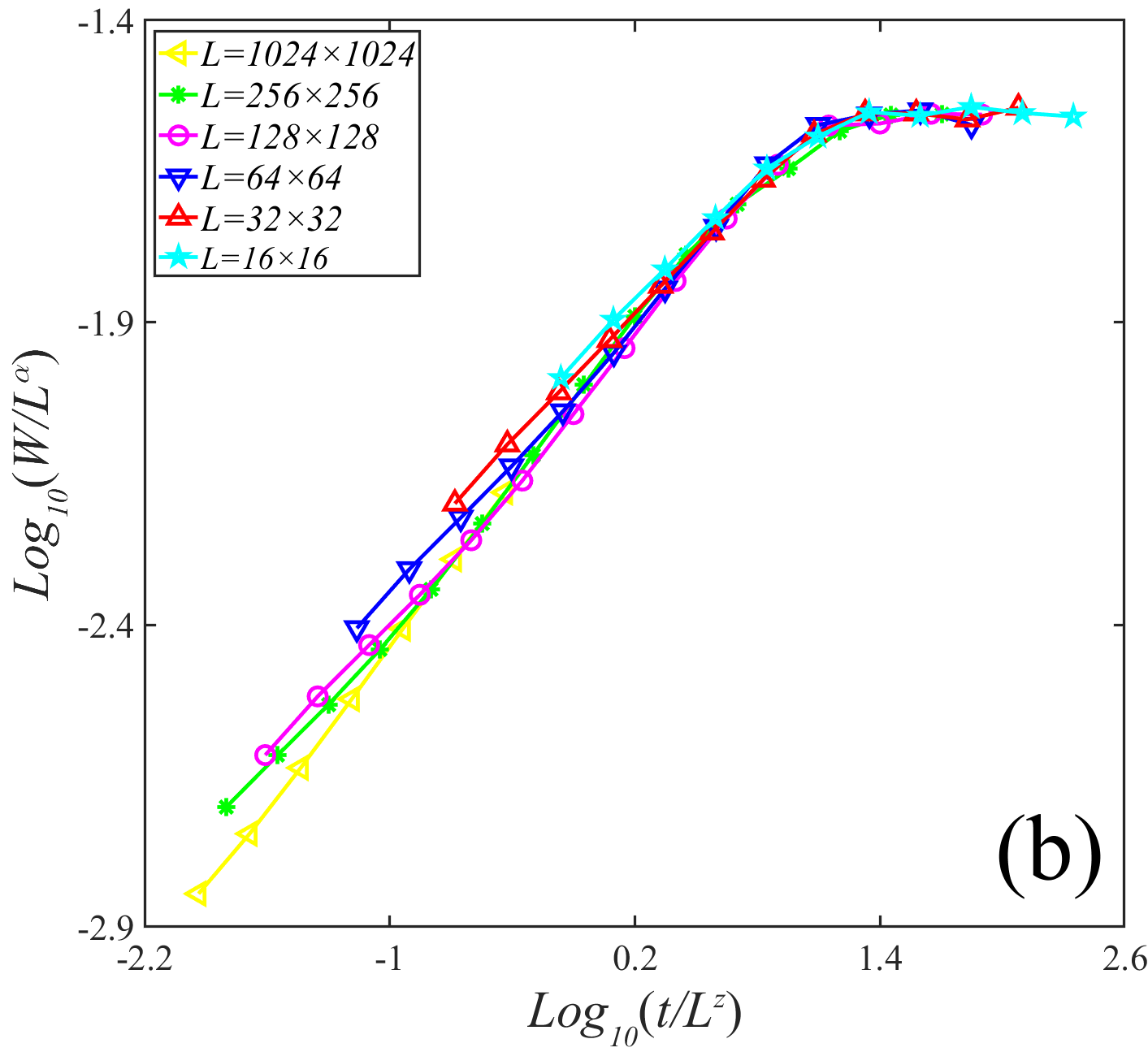}
 
  \caption{ The collapse results of scaled interface width versus the scaled growth time for (2+1)-dimensional KPZ  driven by (a) Gaussian noise with $\alpha=0.385$ and  $z=1.571$, and (b) long-range temporal correlations ($\theta=0.30$) with $\alpha=0.669$ and  $z=1.474$. All data were averaged over 100 independent noise realizations.}
\label{Figure5}
\end{figure}

During training, we first generate data with $L=64\times 64$ and then fine-tuned with $L=256 \times 256$. When $L=64\times 64$, the batch size is set as $N=128$, and the optimizer is Adam. The initial learning rate is $r=0.003$, and if the test error does not decrease for 2 epochs, the learning rate is multiplied by 0.2. When the learning rate is lower than $5\times 10^{-6}$, we fine-tune the model with $L=256 \times 256$. The batch size is changed into $N=32$, the learning rate is $1e-5$. When the learning rate is lower than $5\times 10^{-6}$, we stop the whole training process.

Finally, we perform simulations with the well-trained KPZNet2d. More specifically, we first adopt Gaussian noise to perform simulations on (2+1)-dimensional KPZ, and exhibit the collapse results based on the rescaled $W(L,t)$ and $t$, as shown in Fig.\ref{Figure5}(a). 
Then, we perform simulations  in the presence of temporally correlated noise with $\theta=0.30$ as a typical representative, and exhibit the data collapses on Fig.\ref{Figure5}(b). 
Thus, our results show that the proposed KPZNet2d can perform effectively simulations on (2+1)-dimensional KPZ system without numerical divergence.

\section{Conclusions}
  To summarize, we proposed a numerically stable neural network to simulate the KPZ equation driven by uncorrelated and correlated noises in order to avoid  numerical divergence, and obtain the expected scaling exponents in both (1+1) and (2+1) dimensions. We trained the KPZNet to represent two determined terms of the KPZ equation  by introducing Gaussian white noise to the neural network. The experimental results prove that the KPZNet displays better numerical stability than the commonly used numerical methods. We also adopted the KPZNet to simulate the KPZ equation driven by both long-range temporal and spatial correlations, and the estimated scaling exponents could be achieved effectively.

 This work makes an exploratory attempt in the direction of utilizing the neural network to simulate the stochastic PDEs. The proposed KPZNet is very robust for dealing with the chosen KPZ system behind the training data, and it could be generalized to process interface growth with different distributions. Once the KPZNet is commendably trained, it can be adapted to various random environments eliminating numerical divergence, which is a challenging task for traditional simulation methods.

 In addition to dealing effectively with numerical stability, the neural network operates in a data-driven manner, and then captures the underlying rules within the training data, which can be extended to process actual experimental data from a lot of natural phenomena like burning front in forests or the evolution of desert topography. Even in the absence of precise continuum equations for describing these natural scenes, the proposed neural network can still learn from real-world data and then investigate the inherent dynamic properties. Furthermore, the neural network possesses latent feature spaces and can be applied to various discrete stochastic models belonging to the same universality class, such as the KPZ or EW system. 
In this kind of shared feature spaces, it is possible to study their commonalities and differences, similar to how the Fourier transform allows studying model properties in the frequency domain. Therefore, the neural network-based approaches have better prospects in a large number of nonlinear stochastic PDEs, and extend to more research fields.

\begin{acknowledgments}
We would like to thank Yancheng Wang and Xiongpeng Hu for many useful discussions and the critical reading of the manuscript. This work is supported by Key Academic Discipline Project of China University of Mining and Technology under Grant No. 2022WLXK04.
\end{acknowledgments}

\nocite{*}

\bibliography{myref}

\end{document}